\documentclass[preprint,preprintnumbers,amsmath,amssymb,floatfix,nofootinbib]
{revtex4}
\usepackage{graphicx}
\usepackage{url}

\begin{document}
\preprint{BNL-HET-08/10}
\title{\boldmath{Dark Matter in the Private Higgs Model}}
\author{C.~B.~Jackson}
\email{cbjackson@bnl.gov}
\affiliation{Physics Department, Brookhaven National Laboratory,
Upton, NY 11973-5000, USA}

\date{\today}

\begin{abstract}
The extremely large hierarchy observed in the fermion mass spectrum remains
as one of the most puzzling and unresolved issues in particle physics.
In a recent proposal, however, it was demonstrated that by introducing one Higgs doublet
(or {\it Private Higgs}) per fermion this hierarchy could be made natural by
making the Yukawa couplings between each fermion and its respective Higgs boson
of order unity.  Among the interesting predictions of the Private Higgs 
scenario is a variety of scalars which could be probed at future collider 
experiments and a possible dark matter candidate.
In this paper, we study in some detail the dark matter sector of the Private
Higgs model.  We first calculate the annihilation cross sections of dark matter in this model 
and find that one can easily account for the observed density of 
dark matter in the Universe with relatively natural values of the model's parameters.
Finally, we investigate the possibility of detecting Private Higgs dark matter 
indirectly via the observation of anomalous gamma rays originating from the galactic
halo.  We show that a substantial flux of photons can be produced from the annihilation 
of Private Higgs dark matter such that, if there is considerable clumping of dark matter
in the galactic halo, the flux of these gamma rays could be observed by
ground-based telescope arrays such as VERITAS and HESS. 

\end{abstract}

\maketitle

\section{Introduction}
\label{sec:intro}

One of the most puzzling issues in the Standard Model is the large hierarchy
observed in the masses of fermions.  For example, in the quark
sector alone, the masses of the heaviest (top) and lightest (up) quarks are separated
by nearly five orders of magnitude.  Conversely, if one assumes that all 
fermions receive their mass via interactions with the {\it same} Higgs doublet
(as in the Standard Model (SM)), the large hierarchy of masses observed in 
the fermion sector translates into a large hierarchy in the Yukawa couplings
of the fermions.

Recently, it has been proposed that the hierarchy of fermion masses can be 
made natural by extending the scalar sector of the SM to include one Higgs doublet 
(or {\it Private Higgs} (PH)) per fermion \cite{Porto:2007ed}.  In this scenario, 
all of the Yukawa couplings can be made of ${\cal{O}}(1)$ by tuning parameters 
of the model.  In other words, the vacuum expectation values (vev's) of each respective
PH field can be made to satisfy $v_f \sim m_f$ such that the hierarchy in 
the fermion mass spectrum becomes natural.

The approach to electroweak symmetry breaking (EWSB) in the PH model is quite 
different than those of other multi-Higgs models.  First, one introduces one 
gauge singlet scalar $S_q$ per quark flavor $q$ and uses the vev's of these fields
along with certain interactions between these fields and the various PH fields to 
induce ``negative-mass-squared'' instabilities.  By using different terms in the 
Lagrangian for the top PH and non-top PH fields, one can easily explain the hierarchy 
in vev's by tuning certain parameters of the model.  As a consequence of this approach, the lighter the fermion is, 
the heavier  its associated PH particle must be in order to explain the smallness 
of the respective vev.
In particular, the mass of the PH particle associated with the up quark can
be shown to lie in the $10^2 - 10^3$ TeV range which is definitely beyond the 
reach of the Large Hadron Collider (LHC).  However, there is interesting phenomenology 
originating from the sector of the top and bottom PH fields along with the singlet scalars.  
In this work, we study a scenario where the physical spectrum of this sector contains 
a light SM-like Higgs boson, a heavy scalar Higgs boson, a pair of charged Higgs bosons and a pseudoscalar Higgs boson.  The last three of these arise mainly from the bottom PH field and all
have masses in the $\sim$ TeV range.  In addition to these, there are also two light
scalars which are admixtures of the singlet states associated with the top and bottom
quarks ($S_t$ and $S_b$).  By construction, $S_t$ and $S_b$ are {\it dark} to interactions
with SM gauge fields and fermions.  While  
we will focus mainly on the light scalars in this work, the heavier Higgs 
bosons could be probed at the LHC via production with bottom quarks (since the
Yukawa coupling between the bottom quarks and the bottom PH field is of
order unity)\footnote{Presumably, the PH partner of the $\tau$ could also 
provide interesting phenomenology; however, we will focus on the quark sector
here.}.   

In order to avoid cross-talk between different quarks, the PH model 
contains a set of six discrete symmetries (one for each quark flavor).  Under these 
symmetries, the right-handed
quarks, their respective PH fields and the gauge singlet scalars $S_q$ are all odd, 
while all other SM fields are even.  The existence of these discrete symmetries provides one 
of the most interesting features of the PH scenario which is the possibility
of a dark matter (DM) candidate.  Scalar DM was originally proposed over twenty
years ago in Ref.~\cite{Silveira:1985rk} and has been studied more recently in several
different scenarios including singlet scalar DM \cite{Holz:2001cb, McDonald:1993ex,
Patt:2006fw,Bertolami:2007wb,He:2007tt,Davoudiasl:2004be} and in the so-called
Inert Doublet Model \cite{Ma:2006km,
Barbieri:2006dq,Cirelli:2005uq,Deshpande:1977rw,Majumdar:2006nt,Casas:2006bd,
Calmet:2006hs}.  However, as we will demonstrate, the features of PHDM can be quite
different from previously studied scenarios.

The remainder of the paper is structured as follows.  First, in 
Section~\ref{sec:the-model}, we review the structure of the PH model and demonstrate
how EWSB is realized in this model.  In section Section~\ref{sec:PHDM}, utilizing the 
observations from WMAP \cite{Spergel:2003cb}, we show that the PH model is able to account 
for all of the observed dark matter in the Universe for relatively natural
values of the model's parameters.  In addition, in Section~\ref{sec:indirect-detect}, we 
consider the possibility of 
detecting PHDM via its annihilation into anomalous gamma rays in the galactic
halo.  We show that, with a favorable distribution of DM in the halo, PHDM could
be detected by ground-based telescopes, but is probably beyond the reach of the
space-based GLAST telescope \cite{Morselli:2002nw}.  Finally, in 
Section~\ref{sec:conclusions}, we conclude.

\section{The Model}
\label{sec:the-model}

The main goal of the Private Higgs model is to account for the extremely large
hierarchy observed in the fermion mass spectrum \cite{Porto:2007ed}.  For purposes
of this paper, we will focus on the quark sector.  In contrast
to the SM, where one introduces a single scalar doublet which couples to all
quarks, the PH scenario democratically introduces one Higgs doublet 
$\phi_q$ $(q = u,d,s,c,t,b)$ per quark.  All of the PH fields are assumed to have 
identical $SU(2) \times U(1)$ quantum numbers as the SM Higgs.  In addition to the
PH fields, the scalar sector of the PH model also contains a set of gauge singlet
scalars $S_q$.  In order to avoid {\it cross talk} between quarks of different 
flavors, a set of six discrete symmetries $K_q$ is imposed on the model.  Under the
$K_q$ symmetries, the 
right-handed quarks ($U_q, D_q$) along with the PH fields and $S_q$ are all odd, i.e.:
\begin{equation}
U_q \to -U_q \, (D_q \to -D_q) \,\,\, , \,\,\, \phi_q \to -\phi_q \,\,\, , \,\,\, 
  S_q \to -S_q \,,
\label{eq:Kq-trans}
\end{equation} 
while all other fields are considered even.  The Lagrangian which is symmetric 
under the $K_q$ symmetries is then given by:
\begin{eqnarray}
{\cal{L}} &=& {\cal{L}}_{SM-H} - 
  \sum_q ( Y_D^{PH}\, \overline{Q}_L \phi_D D_q + 
                 Y_U^{PH}\,\overline{Q}_L \tilde{\phi}_U U_q ) \nonumber\\
&&+ \sum_q \biggl[ \partial_\mu S_q \partial^\mu S_q + 
          (D_\mu \phi_q)^\dagger D^\mu \phi_q \biggr] - V(S_q,\phi_q) \,,
\label{eq:PH-Lagrang}
\end{eqnarray}
where $\tilde{\phi}_U = i \sigma_2 \phi_U$, ${\cal{L}}_{SM-H}$ is the SM
Lagrangian without the Higgs terms and $Y_D^{PH},Y_U^{PH}$ are Yukawa matrices.  
The scalar potential $V(S_q,\phi_q)$ takes the form:
\begin{eqnarray}
\label{eq:PH-full-potent}
V(S_q,\phi_q) &=& \sum_q \biggl\{
  \frac{1}{2} M_{S_q}^2 S_q^2 + \frac{\lambda_S^q}{4} S_q^4 +
  \frac{1}{2} M_{\phi_q}^2 \phi_q^\dagger \phi_q + 
  \lambda_q (\phi_q^\dagger \phi_q)^2 - g_{sq}S_q^2 \phi_q^\dagger \phi_q
  \biggr\} \nonumber\\
\nonumber\\
&+& \sum_{q \ne q^\prime} \biggl\{ 
  a_{qq^\prime}^S S_q^2 S_{q^\prime}^2 + 
  \gamma_{qq^\prime} S_q S_{q^\prime} \phi_q^\dagger \phi_{q^\prime} +
  \chi_{qq^\prime} S_{q^\prime}^2 \phi_q^\dagger \phi_q \nonumber\\
\nonumber
&& \,\,\,\,\,\,\,\,\,\,+ \, 
  a_{qq^\prime}  \phi_q^\dagger \phi_{q^\prime} \phi_q^\dagger \phi_{q^\prime} +
  b_{qq^\prime}  \phi_q^\dagger \phi_q \phi_{q^\prime}^\dagger \phi_{q^\prime} +
  c_{qq^\prime}  \phi_q^\dagger \phi_{q^\prime} \phi_{q^\prime}^\dagger \phi_q
  \biggr\} + h.c. \,,
\end{eqnarray}
where, for stability of the potential, $a_{qq^\prime}, b_{qq^\prime}, 
c_{qq^\prime} < 0$.  In our analysis, we will assume these terms are small and
neglect them in the following.

In the PH model, instead of inducing EWSB through the usual ``negative-mass-squared''
approach where $M_{\phi_q}^2 < 0$, one utilizes the vev's of the singlet fields $S_q$ 
and the interactions between the $S_q$'s and the PH fields.  In particular, for the 
top PH one assumes $M_{\phi_t}^2 > 0$ and induces
EWSB through the $g_{st}$ and $\chi_{qt}$ couplings as well as the vev's of the $S_q$ 
fields.  Thus, taking $g_{st},\chi_{qt} > 0$ and
$\frac{1}{2} M_{\phi_t}^2 - g_{st} \langle S_q \rangle^2 - \sum_{q \ne t} 
\chi_{qt} \langle S_q \rangle^2 \equiv \mu_t^2 < 0$, the top PH is forced
to develop a negative-mass-squared instability which, in turn, spontaneously breaks
the $SU(2)_L \times U(1)_Y$ gauge symmetry.  Therefore, in a sense, the top PH plays 
the role of the SM Higgs.  

In general, the PH scenario can contain many new free parameters in addition to
those of the SM.
In order to simplify our analysis in the following sections, we will make a succession
of approximations.  Thus, our results will not probe the full parameter 
space of the PH scenario, but should be viewed as a first step in this direction.
To begin, we follow Ref.~\cite{Porto:2007ed} and assume that $M_{\phi_t}^2 \ll g_{st}v_s^2$ 
which is in accordance with the symmetry breaking pattern discussed above.  We also 
consider the case where $g_{st} \sim \chi_{qt}$ and $a_{qq^\prime}^S \ll 1$.  To give
the $S_q$ fields a vev, one introduces an instability $M_{S_q}^2 < 0$ such that,
under our assumptions, the potential in the $S_q -\phi_t$ sector reduces to:
\begin{equation}
V(S_q,\phi_t) = \frac{\lambda_S^q}{4} \biggl( S_q^2 - \frac{(v_s^q)^2}{2} \biggr)^2 + 
  \lambda_t (\phi_t^\dagger \phi_t)^2 - g_{st} S_q^2 \phi_t^\dagger \phi_t \,,
\end{equation}
where summation over quark flavor $q$ is implicit and, in principle, the quantity 
$v_s^q$ is a bare parameter.  Minimizing this potential, we find the conditions:
\begin{equation}
\frac{\partial V(S_q,\phi_t)}{\partial S_q} \biggl|_{\langle S_q \rangle , 
  \langle \phi_t \rangle} = \lambda_S^q \biggl( \langle S_q \rangle^2 -
  \frac{(v_s^q)^2}{2} \biggr) - 2 g_{st} \langle \phi_t \rangle^2 = 0 \,,
\label{eq:dVdS}
\end{equation}  
and:
\begin{equation}
\frac{\partial V(S_q,\phi_t)}{\partial \phi_t} \biggl|_{\langle S_q \rangle , 
  \langle \phi_t \rangle} = 2 \lambda_t \langle \phi_t \rangle^2 -
  g_{st} \langle S_q \rangle^2 = 0 \,.
\label{eq:dVdphit}
\end{equation}
Solving these equations for the individual vev's we find:
\begin{equation}
\langle S_q \rangle^2 = \frac{(v_s^q)^2}{2} \biggl( \frac{\lambda_S^q \lambda_t}
  {\lambda_S^q \lambda_t - g_{st}^2} \biggr) \equiv \frac{v_s^2}{2} \,,
\label{eq:vs}
\end{equation}
for the vev of $S_q$ and for the top PH vev: 
\begin{equation}
\langle \phi_t \rangle^2 = \frac{(v_s^q)^2}{4} \biggl( 
  \frac{g_{st} \lambda_S^q}{\lambda_S^q \lambda_t - g_{st}^2} \biggr)
  \equiv \frac{v_h^2}{2} \,.
\label{eq:vh}
\end{equation}
Note that, for simplicity, we have identified the individual vev's $v_s^q$ with one
common parameter $v_s$.  Finally, we also note the relationship between $v_s$ and $v_h$:
\begin{equation}
v_h^2 = \frac{g_{st}}{2 \lambda_t} v_s^2 \,.
\label{eq:vs-vh}
\end{equation}

Next, we consider the non-top PH fields which acquire their vev's in a slightly
different manner.  First, as in the case of the top
PH, the mass parameter $M_{\phi_q}^2$ is assumed to be positive. However, for 
the $\phi_q$ fields (where $q \neq t$), one imposes the condition 
$M_{\phi_q}^2 > g_{sq} v_s^2$ in contrast to the case of the top PH.  Then,
vev's for the non-top PH fields are induced through the cubic term 
$\gamma_{q q^\prime}$ and the vev's $v_s$ and $v_h$.  Again, to simplify our 
analysis, we will make some assumptions.  Specifically, we will assume that:
\begin{equation}
M_{\phi_q}^2 \gg g_{sq} v_s^2 \,\,,\,\, \lambda_q \,
\end{equation}
which is consistent with the symmetry breaking pattern discussed above.  Then,
after $S_q$ and $\phi_t$ pick up vev's, the relevant part of the $\phi_q$ 
potential is:
\begin{equation}
\frac{1}{2} M_{\phi_q}^2 \phi_q^\dagger \phi_q - 
  \frac{\gamma_{qt}}{\sqrt{2}} \frac{v_h v_s^2}{2} \phi_q \,.
\label{eq:b-potent}
\end{equation}
Minimizing this potential, the vev's for the non-top Higgs fields are found to be:
\begin{equation}
\langle \phi_q \rangle = \frac{\gamma_{qt}}{\sqrt{2}} 
  \frac{v_h v_s^2}{M_{\phi_q}^2} \equiv v_q \,.
\label{eq:vq}
\end{equation}
Eq.~(\ref{eq:vq}) summarizes the main result of the PH scenario.  By having
the parameter $\gamma_{qt}$ to be small while keeping $M_{\phi_q}^2$ large,
one is able to make {\it all} Yukawa couplings (which are given by $m_q / v_q$) 
of ${\cal{O}}(1)$ without fine-tuning.  As a consequence of this relation,
one can show from Eq.~(\ref{eq:vq}) that the lighter quarks have associated
PH particles in the $10^2 - 10^3$ TeV range which are definitely beyond the reach
of current or future experiments \cite{Porto:2007ed}.  However, the masses from 
the $\phi_t - S_q$ sector can be naturally light (100's GeV), while the bottom PH particle 
can have masses in the TeV range. 

Finally, inserting Eqs.~(\ref{eq:vh}) and (\ref{eq:vq}) into the Lagrangian of 
Eq.~(\ref{eq:PH-Lagrang}), it is easy to show that the $W^\pm$ mass is given in 
the PH model by:
\begin{equation}
m_W^2 = \frac{1}{2} g v_h^2 \biggl[ 1 + \sum_{q \ne t} \biggl(
  \frac{v_q^2}{v_h^2} \biggr) \biggr] \,.
\label{eq:mw2}
\end{equation}
Obviously, the leading term in the sum comes from the bottom PH; however, even in 
this case, the contribution is of order $m_b^2 / m_t^2 \sim 0.001$.  Thus, the 
contributions to EWSB from quarks lighter than the top are negligible and
our statement above that the role of the SM Higgs boson is being played by the top PH
is verified.

\subsection{Mass Eigenstates and Their Interactions}
\label{subsec:mass-estates}

In this section, we study the top-bottom sector of the PH model in some detail.
In particular, we will consider the case where $\lambda_S^t = \lambda_S^b \equiv
\lambda_S$ and $\lambda_S \ll \lambda_S^q$ for $q \ne t, b$.  Thus,
the gauge singlet scalars associated with the lighter quarks become heavy and
effectively decouple from our analysis.  Then, under our assumptions, 
the scalar potential in the top-bottom sector reduces to:
\begin{eqnarray}
\label{eq:Stb-potent}
V(S_q, \phi_t , \phi_b) &=& \frac{\lambda_S}{4} \biggl[
  \biggl( S_t^2 - \frac{v_s^2}{2} \biggr)^2 + 
  \biggl( S_b^2 - \frac{v_s^2}{2} \biggr)^2 \biggr]
  + \lambda_t (\phi_t^\dagger \phi_t)^2 + \frac{1}{2}M_{\phi_b}^2 \phi_b^\dagger \phi_b
\\
&-& a_{tb}^S (S_t^2 S_b^2) - \gamma_{tb} S_t S_b (\phi_b^\dagger \phi_t +
  \phi_t^\dagger \phi_b)
  - g_{st} \biggl[ S_t^2 \phi_t^\dagger \phi_t + S_t^2 \phi_b^\dagger \phi_b +
  S_b^2 \phi_t^\dagger \phi_t \biggr] \,. \nonumber
\end{eqnarray}
To begin, we expand the PH Higgs fields in the usual way:
\begin{eqnarray}
\phi_t ~=~ 
\left(
\begin{array}{c}
\omega^+ \\ 
\frac{1}{\sqrt{2}}(v_h + h_t + i \chi^0)
\end{array}
\right) \,,
\label{eq:phi_t}
\end{eqnarray}
\begin{eqnarray}
\phi_b ~=~ 
\left(
\begin{array}{c}
H^+ \\ 
v_b + H_b + i A_b
\end{array}
\right) \,,
\label{eq:phi_b}
\end{eqnarray}
while the singlet fields are expanded as:
\begin{equation}
S_{t,b} = \frac{1}{\sqrt{2}} (v_s + \sigma_{t,b}) \,.
\label{eq:S}
\end{equation}
In the above expansions, $\omega^\pm$ and $\chi^0$ are assumed to play the roles of the
usual Goldstone bosons which are eaten by the $W^\pm$ and $Z$, while $H^\pm$ and 
$A_b$ are charged and pseudoscalar Higgs bosons, respectively.
Both the $H^\pm$ and $A_b$ will have masses on the order of $M_{\phi_b} \sim$ TeV and 
could provide interesting phenomenology at the LHC.  Note that we are neglecting mixing
between the ``pure Goldstones'' ($\omega^\pm, \chi^0$) and the ``physical Higgs bosons''
($H^\pm, A_b$).  These mixings are typically of order $\frac{\gamma_{tb} v_b^2}
{M_{\phi_b}^2}$ and, thus, are extremely small.

Inserting the expansions of the Higgs fields from Eqs.~(\ref{eq:phi_t}) - (\ref{eq:S})
into (\ref{eq:Stb-potent}), we first extract the mass terms of the Goldstone bosons 
which we require to vanish:
\begin{equation}
m_{\omega^\pm}^2 = m_{\chi^0}^2 = \lambda_t v_h^2 - \frac{1}{2} g_{st} v_s^2 = 0 \,.
\label{eq:GB-masses}
\end{equation}
Note that this equation is in agreement with Eq.~(\ref{eq:vs-vh}).  

Next, we could attempt to diagonalize the full $4 \times 4$ mass matrix in the $(h_t, H_b, \sigma_t, \sigma_b)$ basis.  However, as shown in Ref.~\cite{Porto:2007ed}, for values
of the model parameters that we consider in our analysis most of the mixings between 
the various scalars are negligible.  In particular, the mixing between $h_t$ and $\sigma_q$
is negligible provided:
\begin{equation}
8 g_{st}^3 \ll (2 g_{st} - \lambda_S)^2 \lambda_t \,\,\,\, \mbox{(for $q = t$)}\,,
\end{equation}
and 
\begin{equation}
\gamma_{bt} \biggl(\frac{m_b}{m_t}\biggr) \ll g_{st} \,\,\,\, \mbox{(for $q = b$)} \,.
\end{equation}
Similarly, the mixing between $\sigma_q$ and $H_b$ can be neglected provided:
\begin{equation}
v_h \biggl(\frac{m_b}{m_t}\biggr) \ll v_s\,.
\end{equation}
All of these conditions are satisfied for the parameter choices in our analysis, hence we
choose to neglect the above mixings.  However, it should be noted that the mixing between
$h_t$ and $H_b$ serves to reproduce the small SM coupling between bottom quarks and the SM-like
Higgs boson.  In the following, we identify $h_t$ and $H_b$ with approximate mass eigenstates
$h^0$ and $H^0$ respectively and assume the coupling between $h^0$ and a pair of bottom quarks
takes its SM value. 

Finally, there can be substantial mixing between the two singlet scalars $\sigma_t$ and
$\sigma_b$ via the $a_{tb}^s$ and $\gamma_{tb}$ terms.  Diagonalizing the $2 \times 2$ 
mass matrix in this sector, we find two mass eigenstates ($\Sigma_1$ and $\Sigma_2$) with
mass eigenvalues:
\begin{eqnarray}
\label{eq:mass-Sig1}
m_{\Sigma_1}^2 &=& \frac{1}{2} \biggl(\lambda_S^2 - g_{st}v_h^2 - a_{tb}^s v_s^2 \biggr)
  - \biggl(a_{tb}^s v_s^2 + \frac{\gamma_{tb}}{\sqrt{2}} v_h v_b \biggr) \sin 2 \alpha \,,\\
\nonumber\\
\label{eq:mass-Sig2}
m_{\Sigma_2}^2 &=& \frac{1}{2} \biggl(\lambda_S^2 - g_{st}v_h^2 - a_{tb}^s v_s^2 \biggr)
  + \biggl(a_{tb}^s v_s^2 + \frac{\gamma_{tb}}{\sqrt{2}} v_h v_b \biggr) \sin 2 \alpha \,.
\end{eqnarray}
Note that for $a_{tb}^s, \gamma_{tb} > 0$ and $0 < \alpha < \pi/2$, $\Sigma_1$ plays the
role of the lightest PH particle (LPHP) which is stable against decay and, thus, provides
a candidate for DM.  Using Eqs.~(\ref{eq:mass-Sig1}) and (\ref{eq:mass-Sig2}), 
we can exchange two of the free parameters (e.g., $\lambda_S$ and $a_{tb}^s$) for the
masses of the two singlet scalars.  This is the approach we will take.  Therefore,
in the analysis to follow, we will take as our free parameters the masses $m_{\Sigma_1}$,
$m_{\Sigma_2}$ as well as the couplings $g_{st}$ and $\gamma_{tb}$ and the mixing angle 
$\alpha$.  Note that
the conditions for small mixings between the $\sigma_q$'s and the PH fields forces 
$g_{st}$ to take small values.

\section{Private Higgs Dark Matter}
\label{sec:PHDM}

As mentioned earlier, one of the most interesting aspects of the PH 
scenario is the prospect of a Weakly Interacting Massive Particle (WIMP)
with masses in the expected natural range for DM.  In this context, the PH
model is similar to other scalar DM models such as the gauge singlet models of
Refs.~\cite{Silveira:1985rk,Holz:2001cb, McDonald:1993ex,
Patt:2006fw,Bertolami:2007wb,He:2007tt,Davoudiasl:2004be} and the Inert Doublet Model (IDM) 
\cite{Ma:2006km,Barbieri:2006dq,Cirelli:2005uq,Deshpande:1977rw,Majumdar:2006nt,
Casas:2006bd,Calmet:2006hs}.  In this section,
we calculate the annihilation cross sections of PHDM into SM particles and show that,
for relatively natural values of the model parameters, one can account
for all of the observed dark matter in the Universe.  In the next section, we
investigate the possibility of indirectly detecting PHDM via its 
annihilation into {\it anomalous} gamma rays in the galactic halo.

\begin{figure}[t]
\begin{center}
\includegraphics[scale=0.5]{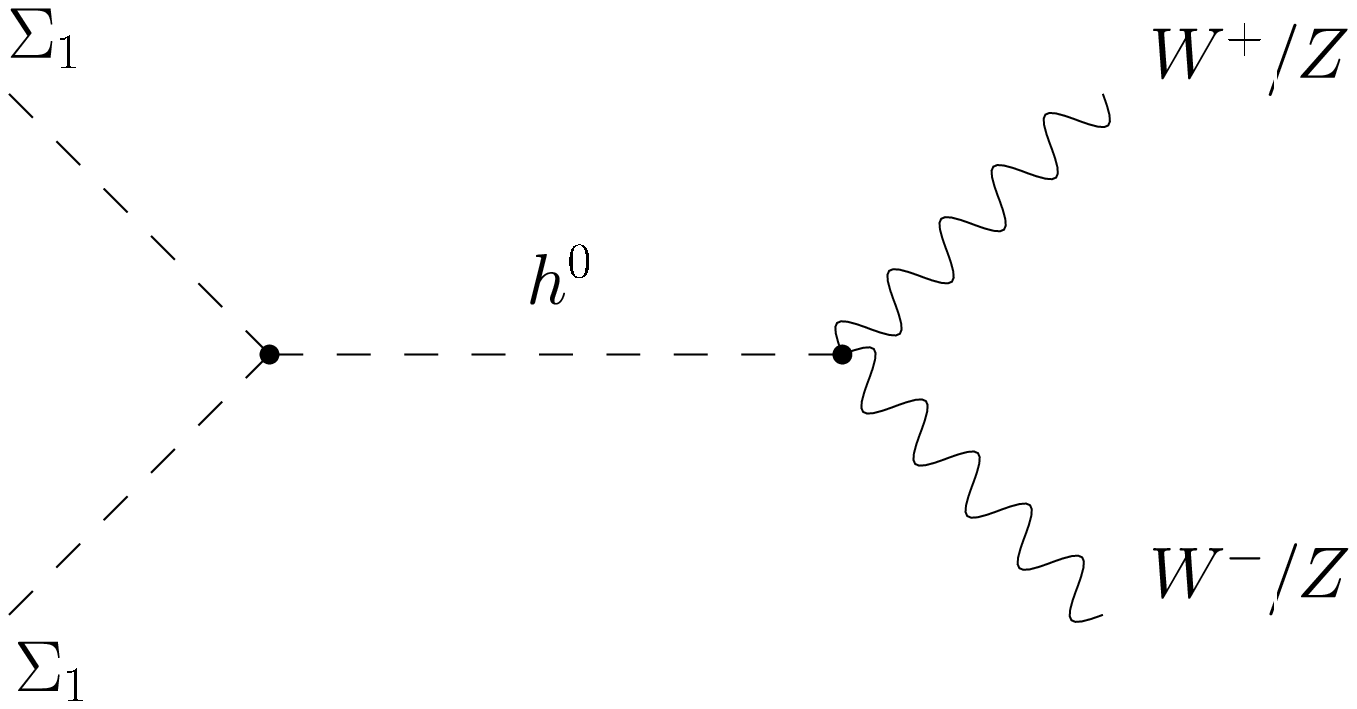}
\hspace{0.5cm}
\includegraphics[scale=0.5]{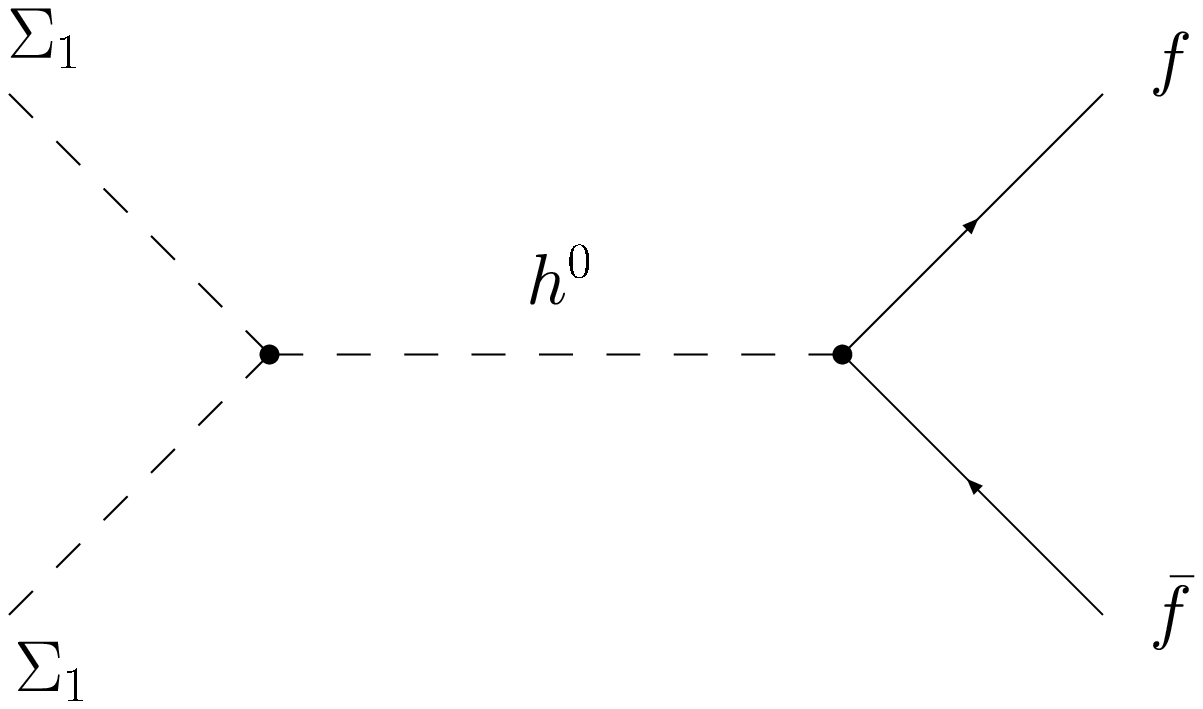}
\end{center}
\caption[]{Leading $s$-channel processes which maintain the singlet scalar $\Sigma_1$ in
  equilibrium with the rest of the cosmic fluid.}
\label{fg:SigSig-ann-schan}
\end{figure}

First, let us consider the present relic abundance of PHDM in the Universe. 
In the following, we will assume that the mass splitting $m_{\Sigma_2}-m_{\Sigma_1}$
is large enough that coannihilation reactions between $\Sigma_1$ and $\Sigma_2$
do not significantly affect the relic abundance.  These effects will be considered
in future work.
In the early Universe, the singlet scalar $\Sigma_1$ would have been in 
equilibrium with the rest of the cosmic fluid.  This equilibrium is maintained
via $\Sigma_1$ pair-annihilation and pair-creation reactions which proceed through
the $s$-channel exchange of the SM-like Higgs $h^0$.  The leading 
$2 \to 2$ $s$-channel reactions which contribute to these processes are shown in Fig.
\ref{fg:SigSig-ann-schan}.

The present relic abundance of PHDM is determined by the pair-annihilation
rates in the non-relativistic limit.  The rates for each allowed channel are 
given in the non-relativistic limit as:
\begin{equation}
a(X) \equiv \lim_{u \to 0} \sigma(\Sigma_1 \Sigma_1 \to X) \, u\,
\label{eq:aX}
\end{equation}
where $u$ is the relative velocity of the annihilating particles.  The total
annihilation cross section is then given by summing over each of the allowed 
channels.  Computing the cross sections for the diagrams in 
Fig.~\ref{fg:SigSig-ann-schan}, we find:
\begin{eqnarray}
\label{eq:aWW}
a(W^+ W^-) &=& \frac{g_{\Sigma_1 \Sigma_1 h^0}^2}{2\pi v_h^2} \frac{\sqrt{1 - \mu_w}}
  {(4m_{\Sigma_1}^2 - m_{h^0}^2)^2 + \Gamma_{h^0}^2 m_{h^0}^2} 
  \biggl( 1 - \mu_w + \frac{3}{4} \mu_w^2 \biggr) \, , \\
\nonumber\\
a(ZZ) &=& \frac{g_{\Sigma_1 \Sigma_1 h^0}^2}{4\pi v_h^2} \frac{\sqrt{1 - \mu_z}}
  {(4m_{\Sigma_1}^2 - m_{h^0}^2)^2 + \Gamma_{h^0}^2 m_{h^0}^2} 
  \biggl( 1 - \mu_z + \frac{3}{4} \mu_z^2 \biggr) \, , \\
\nonumber\\
a(f\bar{f}) &=& \frac{g_{\Sigma_1 \Sigma_1 h^0}^2}{4\pi v_h^2} 
  \frac{(1 - \mu_f)^{\frac{3}{2}}}
  {(4m_{\Sigma_1}^2 - m_{h^0}^2)^2 + \Gamma_{h^0}^2 m_{h^0}^2} \, ,
\end{eqnarray}
where $\mu_i = m_i^2 / m_{\Sigma_1}^2$, $\Gamma_{h^0}$ is the width of the $h^0$
for which we use SM values and the expression for the coupling 
$g_{\Sigma_1 \Sigma_1 h^0}$ is given by:
\begin{equation}
\label{eq:coupSigSigh}
g_{\Sigma_1 \Sigma_1 h^0} = -g_{st} v_h + \frac{2 \gamma_{tb} v_b}{\sqrt{2}}
  c_\alpha s_\alpha \,.
\end{equation}

The WMAP collaboration \cite{Spergel:2003cb} provides a very precise determination 
of the present DM abundance which, at the two-sigma level, is given by:
\begin{equation}
\Omega_{DM} h^2 = 0.111 \pm 0.018 \,.
\end{equation}
As shown in Ref.~\cite{Birkedal:2004xn}, for a generic model of DM, the present 
abundance of DM is mainly determined by $J_0$ (the angular momentum of the dominant 
partial wave contributing
to DM annihilation) and the annihilation cross section.  In contrast, the relic 
abundance depends only weakly on the mass or spin of the DM particle.  Thus, the
very precise constraints from WMAP on $\Omega_{DM} h^2$ translate into very 
precise constraints on the quantity $a \equiv \sum_X a(X)$ depending on the 
value of $J_0$.  In particular, for an ``$s$-wave annihilator'' ($J_0 = 0$) such as
the case considered here, the WMAP measurement translates into the bounds:
\begin{equation}
a = 0.8 \pm 0.1 \,\,\mbox{pb} \,,
\label{eq:a-bounds}
\end{equation}
nearly independent of the mass or spin of the DM particle (see Fig. 1 of 
Ref.~\cite{Birkedal:2004xn}).

\begin{figure}[t]
\begin{center}
\includegraphics[scale=0.5]{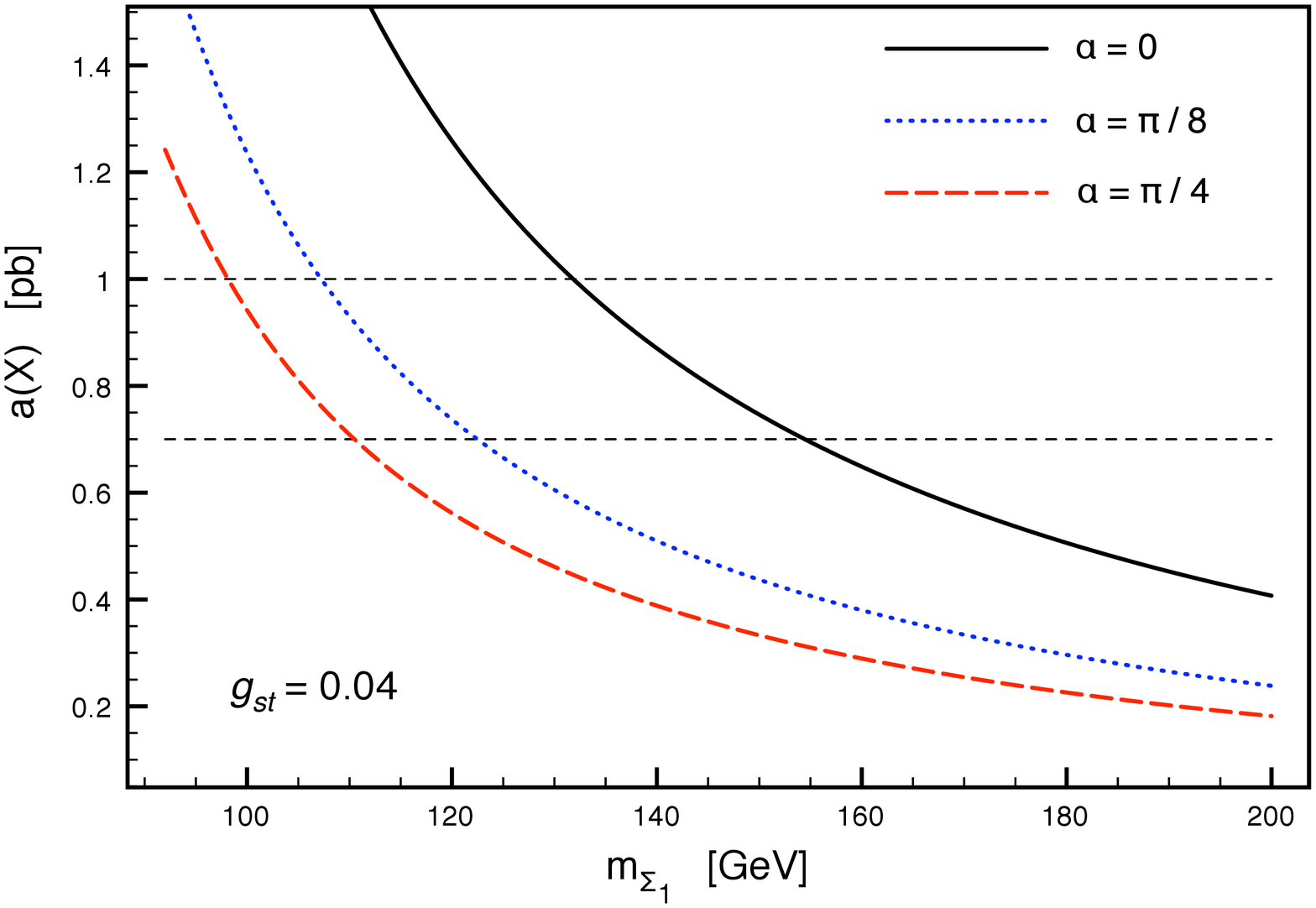} \\
\vspace{1cm}
\includegraphics[scale=0.5]{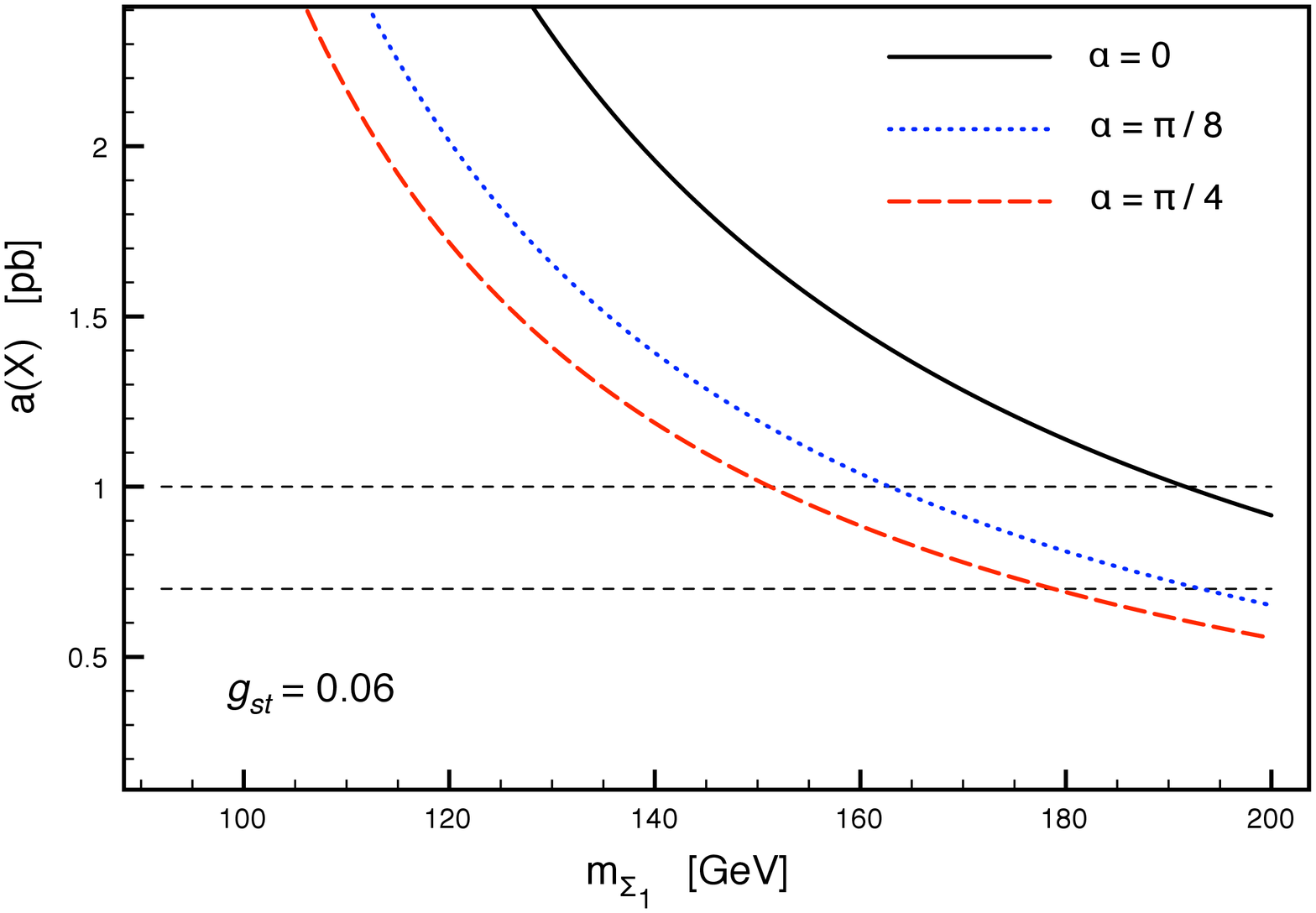}
\end{center}
\caption[]{The annihilation cross section as a function of the $\Sigma_1$ mass and 
  the mixing parameter $\alpha$.  The dashed horizontal lines indicate the WMAP
  constraints on the annihilation cross sections given by Eq.~\ref{eq:a-bounds}.}
\label{fg:SigSig-ann}
\end{figure}

In Fig.~\ref{fg:SigSig-ann}, we plot the values of $a(X)$ for two different values
of the coupling $g_{st}$ and several different values of the mixing angle $\alpha$.  
In these plots, we have set the bottom PH vev equal to the 
bottom quark mass and $\gamma_{tb} = 1$.  The horizontal dashed lines indicate the limits
on $a(X)$ from Eq.~(\ref{eq:a-bounds}).  Clearly, from Eqs.~(\ref{eq:aWW})-
(\ref{eq:coupSigSigh}), we see that the annihilation cross sections depend quadratically
on $g_{st}$.  This is evident in the plot of Fig.~\ref{fg:SigSig-ann} where we see
a small shift in $g_{st}$ results in a large shift in the ranges of $m_{\Sigma_1}$ 
allowed by the WMAP data.  Finally, we note that these plots are only meant 
to show that for relatively natural values of the model parameters it is indeed possible to 
account for the observed density of DM in the Universe.  A full scan of the PH 
parameter space would probably find other choices of parameters which could 
fulfill the constraints from Eq.~(\ref{eq:a-bounds}).

\section{Indirect Detection of PHDM}
\label{sec:indirect-detect}

Next, we would like to investigate the possibility of detecting PHDM.  We will
focus here on indirect detection and save an analysis of direct detection for
future work.

As we have seen, the annihilation rates for PHDM are approximately 
velocity-independent in the non-relativistic regime.  In general, this implies that DM
collected in galactic halos has a substantial probability to pair-annihilate
resulting in anomalous high-energy cosmic rays which can be distinguished 
from astrophysical backgrounds.  In particular, gamma rays from these annihilations
provide a chance to extract information about DM, since they can travel over 
galactic scales without scattering.

\begin{figure}[t]
\begin{center}
\includegraphics[scale=0.4]{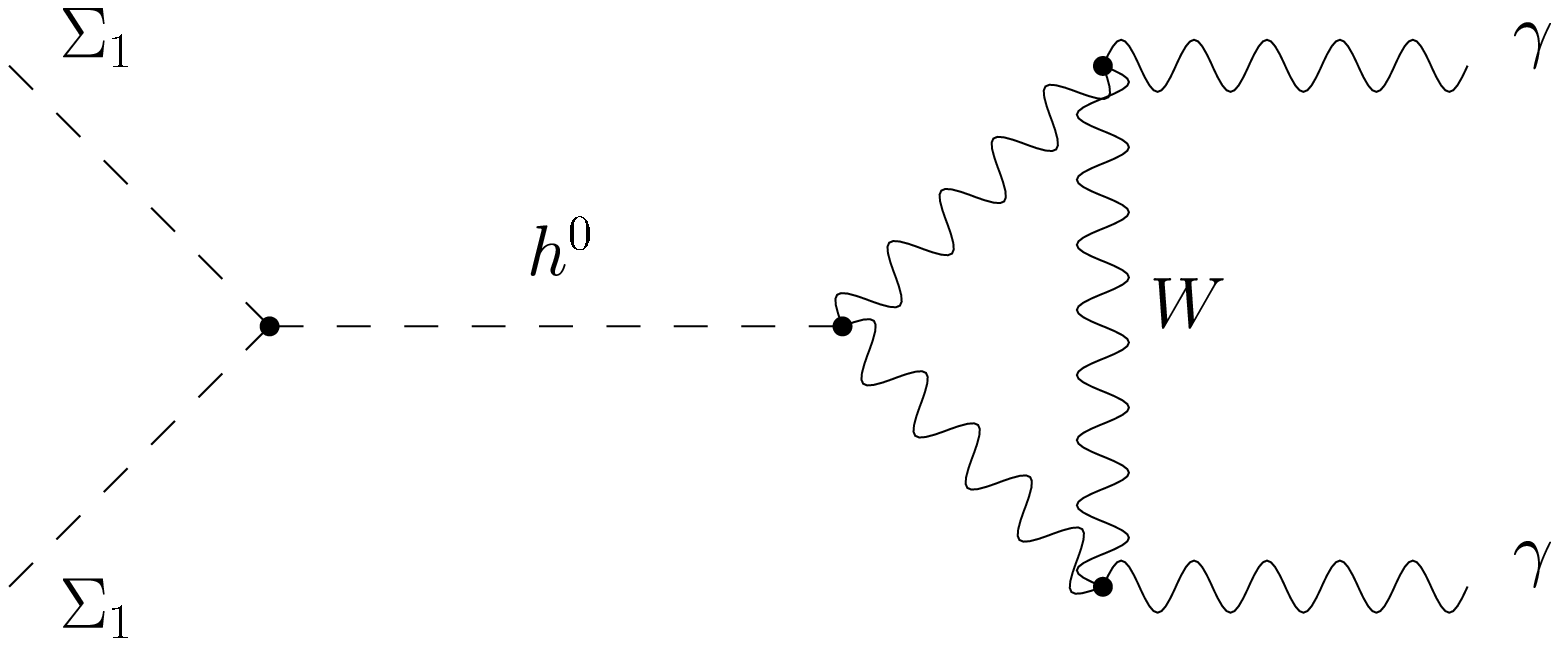}
\hspace{0.5cm}
\includegraphics[scale=0.4]{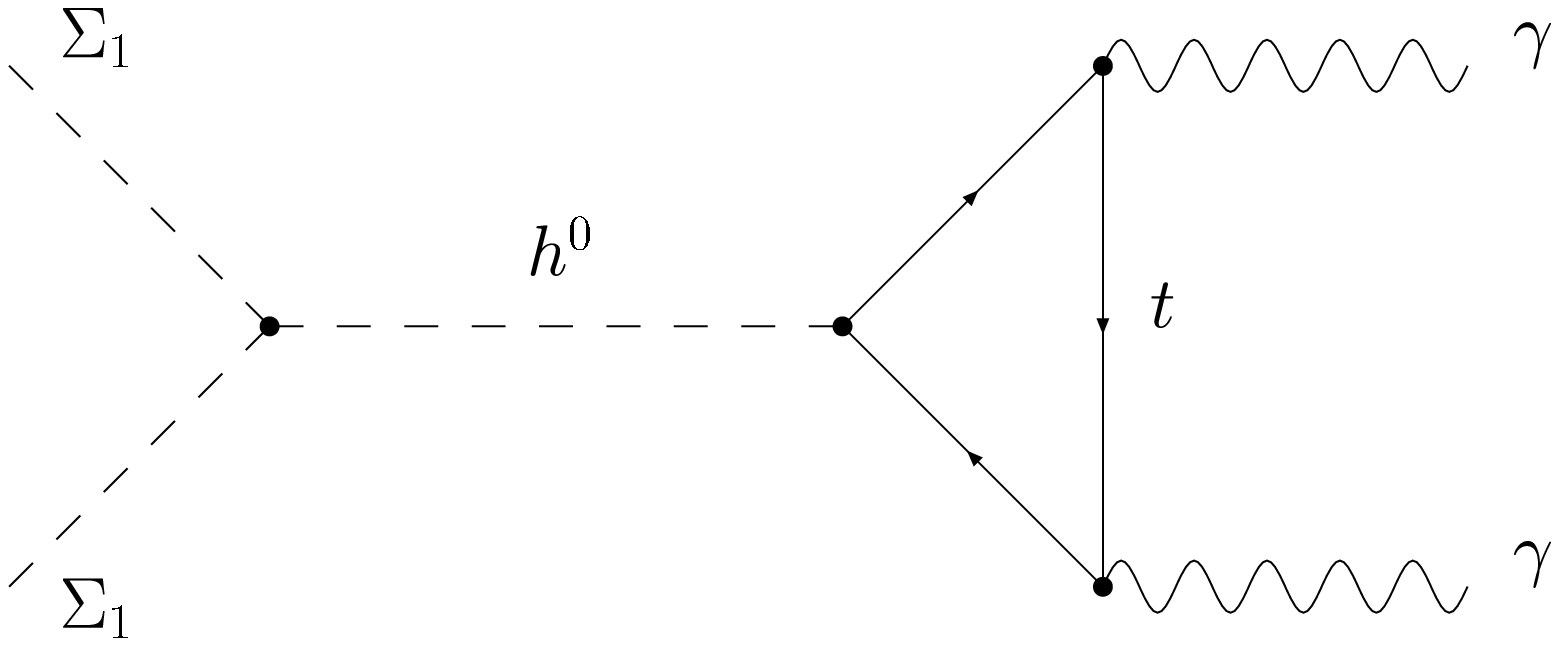}
\end{center}
\caption[]{Diagrams which dominate photon pair-production in the $\Sigma_1$ 
  annihilation in the galactic halo.}
\label{fg:SigSig-to-phopho-loops}
\end{figure}

The production of gamma rays from $\Sigma_1 \Sigma_1$ annihilation can originate from 
several different processes (including hadronization, factorization and 
radiation from final-state particles).  However, for simplicity, we will assume
the dominate source is from direct annihilation into a two-body final state 
as shown in Fig.~\ref{fg:SigSig-to-phopho-loops}\footnote{Here, we concentrate on the
dominant $\gamma\gamma$ signal and save a discussion of the $Z\gamma$ and/or
$h^0\gamma$ channels for future work.}.  Note that, under our assumptions,
only SM particles circulate the loop.  In the full parameter space of the PH
model, it would be possible to have charged Higgs circulating the loop.  However,
their couplings to $h^0$ are {\it always} of order $v_b / v_h$ or $v_b / v_s$
and, thus, can be safely ignored in comparison to the SM loops.  The cross section 
for photon pair-production in the PH scenario can be written as:
\begin{equation}
\sigma_{\gamma\gamma} u = \frac{2 g_{\Sigma_1 \Sigma_1 h^0}^2}
  {(s - m_{h^0}^2)^2 + \Gamma_{h^0} m_{h^0}^2} 
  \frac{\hat{\Gamma}(h^0 \to \gamma\gamma)}{\sqrt{s}} \,,
\label{sec:phopho-cs}
\end{equation}
where the expression for $g_{\Sigma_1 \Sigma_1 h^0}$ is given above and, in the 
non-relativistic regime, $s \simeq 4 m_{\Sigma_1}^2$.  The hat on
${\hat{\Gamma}}$ indicates that one should replace $m_{h^0} \to \sqrt{s}$ in the 
standard expressions for on-shell Higgs decays.  The expressions needed to 
construct $\hat{\Gamma}(h^0 \to \gamma\gamma)$ can be found in several reviews
(e.g., see Ref.~\cite{Djouadi:2005gi}) and, hence, we will not repeat them here.

Next, we would like to compute the flux of photons observed on Earth from 
$\Sigma_1$ annihilation in the galactic halo.  The monochromatic flux due to the
$\gamma\gamma$ final state, observed by a telescope with a line of sight 
parameterized by $\Psi = (\theta, \phi)$ and a field of view $\Delta \Omega$
can be written as \cite{Bergstrom:1997fj}:
\begin{equation}
\Phi = (1.1 \times 10^{-9} \,\mbox{s}^{-1} \mbox{cm}^{-2}) 
  \biggl( \frac{\sigma_{\gamma\gamma}u}{1\,\,\mbox{pb}}\biggr)
  \biggl( \frac{100 \,\,\mbox{GeV}}{m_{\Sigma_1}} \biggr) 
  \bar{J}(\Psi, \Delta\Omega) \Delta\Omega \,,
\label{eq:flux}
\end{equation}
where the dependence of the flux on the halo dark matter density distribution is
contained in $\bar{J}$.  Many models predict a large spike in the DM density 
in the neighborhood of the galactic center, making the line of sight towards
the center of the galaxy the preferred one.  However, the features of the peak
are highly model-dependent resulting in values of $\bar{J}$ ranging from 
$10^3$ to $10^7$ for $\Delta\Omega = 10^{-3}$ sr (typical for ground-based
atmospheric Cerenkov telescopes) \cite{Navarro:1996gj,Moore:1997sg,
Moore:1999gc,Gnedin:2004cx}. 

\begin{figure}[t]
\begin{center}
\includegraphics[scale=0.5]{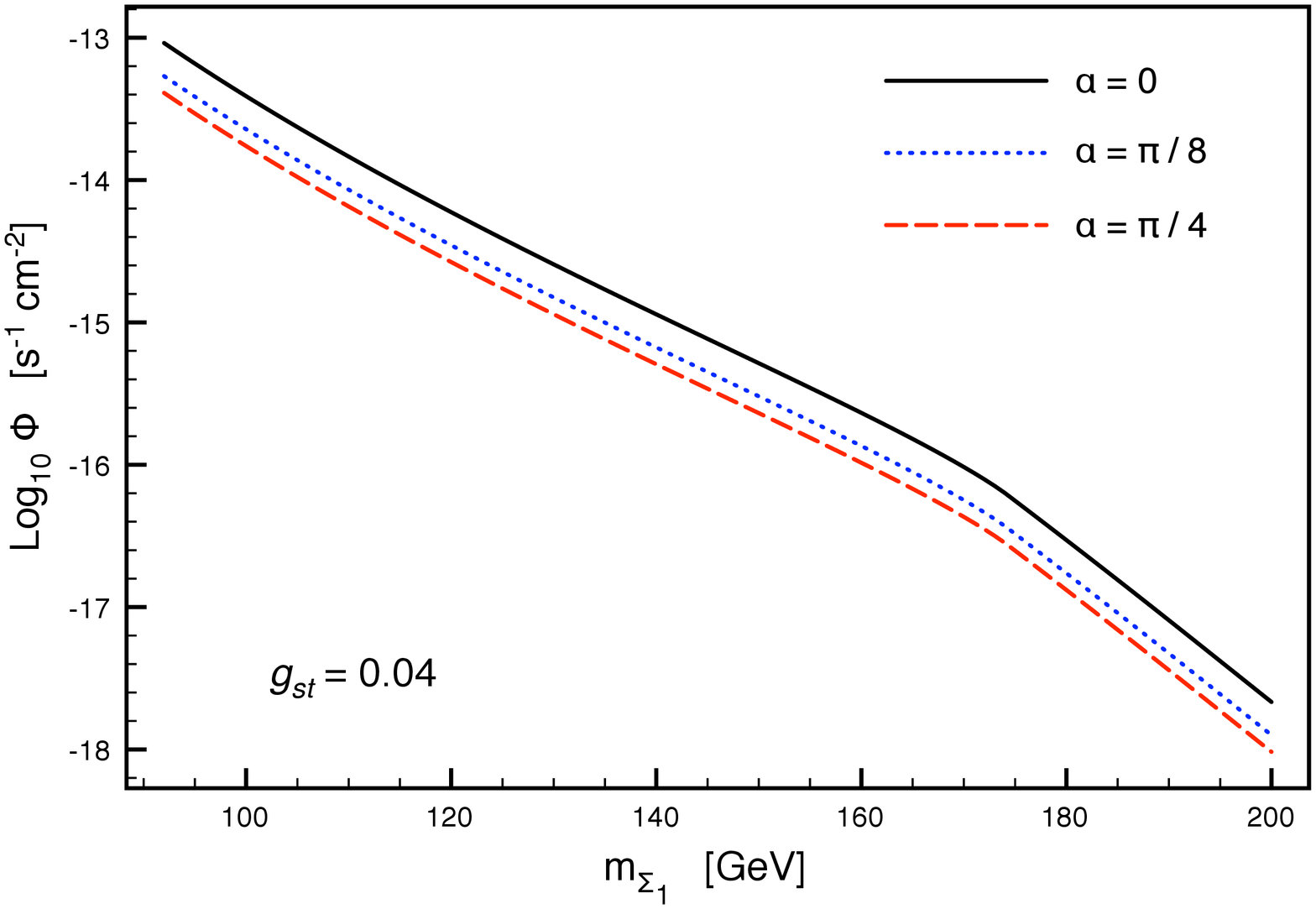} \\
\vspace{1cm}
\includegraphics[scale=0.5]{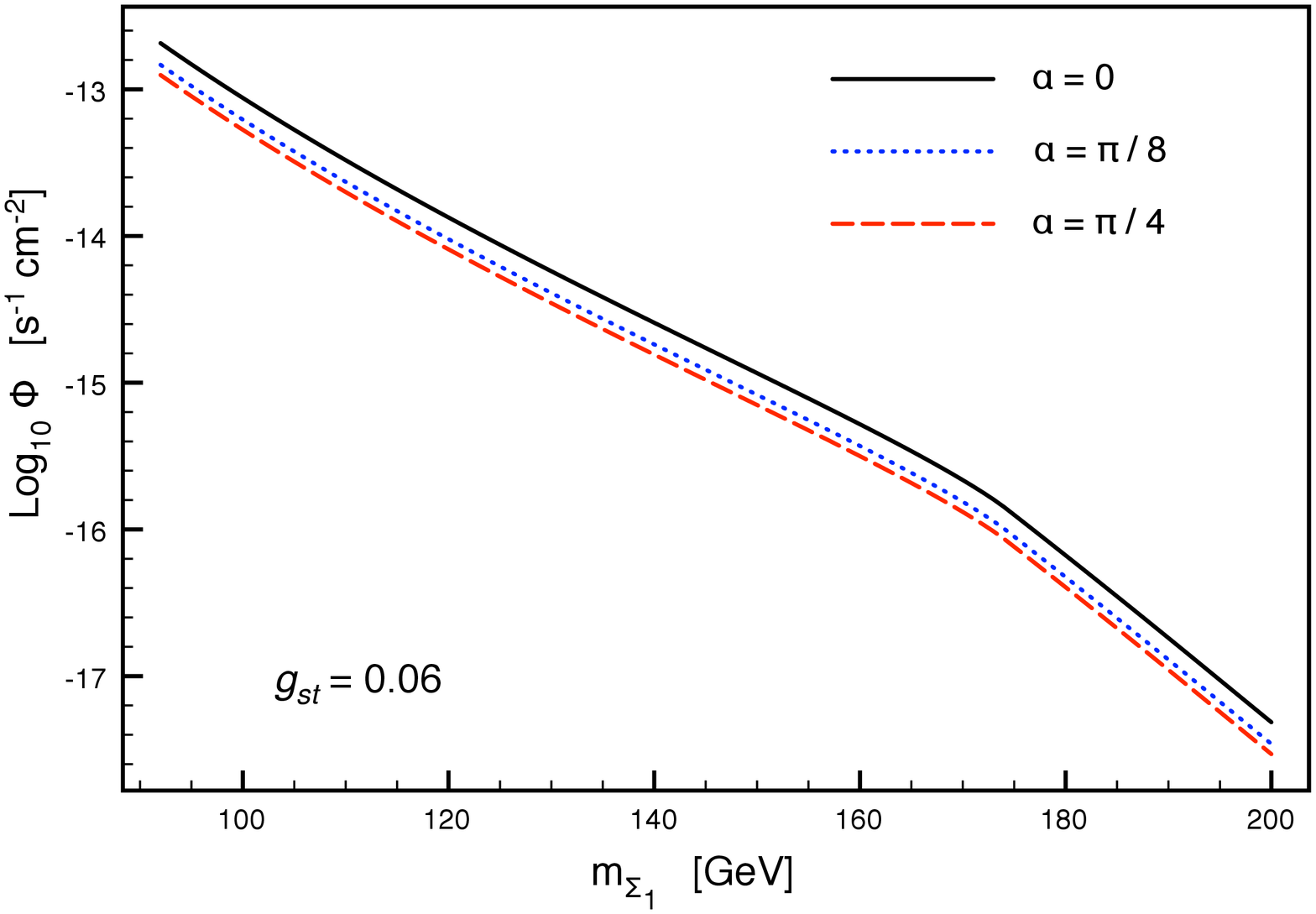}
\end{center}
\caption[]{The flux of monochromatic photons from the reaction $\Sigma_1 \Sigma_1 
  \to \gamma\gamma$ for $\bar{J}\Delta\Omega = 1$ for two different values 
  of the coupling $g_{st}$.}
\label{fg:PhotonFlux-genmix}
\end{figure}

The monochromatic photon flux predicted from PHDM annihilation for the two values
of $g_{st}$ studied previously are shown in Fig.~\ref{fg:PhotonFlux-genmix}.
For these plots, we have assumed there is no 
substantial spiking in the galactic center (i.e., $\bar{J}\Delta\Omega = 1$).  
In the energy range considered in these plots, 
ground-based 
atmospheric Cherenkov telescopes (such as VERITAS \cite{Weekes:2001pd} and HESS 
\cite{Hinton:2004eu}) typically have a flux sensitivity down to the $10^{-12}$
s$^{-1}$ cm$^{-2}$ level.
On the other hand, the upcoming space-based telescope
GLAST \cite{Morselli:2002nw} is limited by statistics to the  $10^{-10}$ 
s$^{-1}$ cm$^{-2}$ level over the energy range considered.
From these plots, it is clear that without a substantial spike in the galactic 
center, PHDM will be difficult to observe in either ground- or space-based observatories.
However, if the halo does exhibit a substantial spike or strong clumping
(e.g., if $\bar{J} \ge 10^5$ at $\Delta\Omega \simeq 10^-3$), PHDM could 
be observed at ground-based telescopes assuming small values of $g_{st}$ and 
relatively light masses ($m_{\Sigma_1} \simeq 100 - 120$ GeV).

\section{Conclusions}
\label{sec:conclusions}

The Private Higgs model attempts to address the large hierarchy observed in the fermion
mass spectrum by introducing one Higgs doublet for each fermion.  EWSB is achieved
not by the usual ``negative-mass-squared'' approach, but by introducing a set of gauge
singlet scalars and using the vev's of these fields and their interactions with the PH
fields to induce instabilities.  In order to avoid cross-talk between quarks of
different flavors, one
also introduces a set of discrete symmetries.  
This provides one of the most interesting features of the 
Private Higgs model: a possible dark matter candidate.

In this paper, we have begun an investigation of the PH dark matter sector.  
We found that for relatively natural
values of the model's parameters the PH model provides a candidate which 
can account for the relic density of dark matter observed in the present Universe.
To show this, we calculated the annihilation cross section for PHDM into SM 
particles and compared to limits on the cross section which can be obtained from 
the WMAP observations.  

Finally, we investigated the possibility of detecting PHDM via anomalous gamma rays 
originating from the annihilation of PHDM in the galactic halo.  While the observation of
these gamma rays may be difficult for the space-based GLAST observatory, we showed 
that evidence of PHDM could be observed at ground-based atmospheric Cerenkov telescopes
such as VERITAS and HESS if there is substantial clustering of dark matter in the 
galactic halo.

\begin{acknowledgements}
The author is very grateful to Sally Dawson for a careful reading of this
manuscript and Rafael Porto for extremely useful discussions on the Private Higgs
model.  This manuscript has 
been authored by employees of Brookhaven Science Associates, LLC under 
Contract No. DE-AC02-98CH10886 with the U.S. Department of Energy. The 
publisher by accepting the manuscript for publication acknowledges that the 
United States Government retains a non-exclusive, paid-up, irrevocable, 
world-wide license to publish or reproduce the published form of this 
manuscript, or allow others to do so, for United States Government purposes.
\end{acknowledgements}

\bibliography{PH-paper-revised}

\end{document}